\documentclass{aa}

\usepackage{graphicx}
\usepackage{txfonts}
\usepackage{tabularx}
\usepackage{rotating}
\usepackage{booktabs}
\usepackage{multirow}
\usepackage[switch]{lineno}
\usepackage{graphicx}

\modulolinenumbers[5]

\begin{document}
	
\title{Planets Around Solar Twins/Analogs (PASTA) III: Chemical Clock Relations in Planet-Hosting Solar Analogs}

\authorrunning{Sun et al.}

\author{Qinghui Sun\inst{1}\thanks{Corresponding author: \email{qinghuisun@sjtu.edu.cn}}
	\and Serat M. Saad\inst{2}
	\and Yuan-Sen Ting\inst{2,3,4}
	\and Jiayue Zhang\inst{5}
	\and Chenyang Ji\inst{5}
	\and Fei Dai\inst{6}
	\and Sharon Xuesong Wang\inst{5}
	\and Bryant Randolph\inst{1}
	\and Yunzhe Gu\inst{1}
	\and Ilya Ilyin\inst{7}
	\and Yaguang Li\inst{6}
	\and Jie Yu\inst{8,9,10}
	\and Fan Liu\inst{11}
}

\institute{
	Tsung-Dao Lee Institute, School of Physics and Astronomy, \& State Key Laboratory of Dark Matter Physics, Shanghai Jiao Tong University, Shanghai 201210, China\\
	\email{qinghuisun@sjtu.edu.cn}
	\and
	Department of Astronomy, The Ohio State University, Columbus, OH 43210, USA
	\and
	Center for Cosmology and AstroParticle Physics (CCAPP), The Ohio State University, Columbus, OH 43210, USA
	\and
	Max-Planck-Institut f${\ddot u}$r Astronomie, K${\ddot o}$nigstuhl 17, D-69117 Heidelberg, Germany
	\and
	Department of Astronomy, Tsinghua University, Beijing, 100084, China
	\and
	Institute for Astronomy, University of Hawai‘i, 2680 Woodlawn Drive, Honolulu, HI 96822, USA
	\and
	Leibniz-Institut for Astrophysics Potsdam (AIP), An der Sternwarte 16, D14482 Potsdam, Germany
	\and 
	School of Astronomy and Space Science, Nanjing University, Nanjing 210023, People's Republic of China
	\and
	Key Laboratory of Modern Astronomy and Astrophysics, Ministry of Education, Nanjing 210023, People's Republic of China
	\and
	Research School of Astronomy \& Astrophysics, Australian National University, Cotter Rd., Weston, ACT 2611, Australia
	\and
	National Astronomical Observatories, Chinese Academy of Sciences, Beijing 100101, China
}
	
\abstract
{Abundance ratios between elements synthesized on different nucleosynthetic timescales provide empirical chemical clocks for estimating stellar ages, but robust calibrations require homogeneous samples with precise stellar parameters and elemental abundances.}
{We aim to investigate whether the presence of planets modifies empirical chemical clock relations.}
{We present high-precision stellar parameters and elemental abundances for 52 new planet-hosting targets observed with Keck/HIRES, LBT/PEPSI, and Subaru/HDS. Combined with PASTA I and II, the full sample comprises 94 stars, of which 81 are identified as planet-hosting solar twins and analogs, and is used to characterize Galactic chemical evolution and derive empirical chemical clock relations.}
{We recover the expected Galactic chemical evolution trends, with $\alpha$-elements such as Mg and S increasing with stellar age, while $s$-process elements such as Y, Ba, and Sr show negative age correlations. In contrast, Fe-peak elements and the $r$-process element Eu show little or no significant age dependence. These trends give rise to tight chemical clock relations based on [Y/Mg], [Y/Al], [Ba/Mg], [Ba/Al], and [Sr/Mg], with [Y/Mg] and [Y/Al] providing the strongest correlations. [Ba/Al] shows the steepest age dependence, and the Ba-based ratios are generally more sensitive to age than their Y-based counterparts.}
% conclusion
{Planet-hosting solar analogs follow chemical-clock relations consistent with those of nearby solar twins, with no evidence for a significant planet-related effect. The derived relations are primarily applicable to local solar twins and analogs in the Galactic thin disk.}

\maketitle
	
\section{Introduction}

The chemical composition of stars encodes the nucleosynthetic history of the Galaxy and provides a powerful tool for studying Galactic chemical evolution (GCE; e.g., \citealt{2009ARA&A..47..481A, 2016ARA&A..54..529B}). However, abundance patterns observed in stars of different types can be affected by stellar evolutionary processes and systematic uncertainties in spectroscopic analyses, which introduce additional scatter and can obscure subtle Galactic trends. In this context, solar twins and analogs offer a particularly advantageous laboratory, as their similar stellar parameters enable strictly differential analyses relative to the Sun, thereby minimizing systematic effects and allowing GCE to be probed with reduced contamination from stellar evolutionary effects \citep[e.g.,][]{2009ApJ...704L..66M}. Over the past decade, high-precision spectroscopic studies of solar twins have demonstrated that elemental abundances can be measured with a precision of $\sim$0.01~dex through strict line-by-line differential analyses relative to the Sun, enabling detailed investigations of chemical trends in the Galactic disk \citep[e.g.,][]{2018ApJ...865...68B, 2025ApJ...980..179S, 2025A&A...701A.107S, 2025A&A...701A.153S}. High-precision studies have shown that a large fraction of the observed abundance differences among solar-type stars can be explained by Galactic chemical evolution (GCE) effects \citep[e.g.,][]{2016AA...593A.125S, 2018ApJ...865...68B}, highlighting the ability of solar twins and analogs to trace subtle GCE trends.

A key result from these studies is that many elemental abundance ratios exhibit tight correlations with stellar age. In particular, elements produced on different nucleosynthetic timescales trace the time evolution of the interstellar medium \citep[e.g.,][]{2001ApJ...547..217T}. $\alpha$-elements such as Mg and S are predominantly produced by massive stars and enrich the interstellar medium on short timescales, whereas neutron-capture elements such as Y and Sr are largely produced by the $s$-process in asymptotic giant branch stars on longer timescales. We note that Al is an odd-$Z$ light element rather than an $\alpha$-element in the usual nucleosynthetic classification, although its Galactic abundance evolution and age dependence can resemble those of $\alpha$-elements. Consequently, abundance ratios combining elements with different enrichment timescales, such as [Y/Mg] and [Y/Al], show strong, nearly linear correlations with stellar age in solar twins \citep{2012MNRAS.421.1231T, 2015AA...579A..52N, 2016AA...593A.125S}. These relations form the basis of so-called ``chemical clocks,'' which provide an empirical method for estimating stellar ages from chemical abundances.

Chemical clocks have become an important tool in Galactic archaeology, providing age estimates with precisions approaching $\sim$1~Gyr for solar twins. However, subsequent work has shown that these relations are not universal. The calibration of abundance--age relations depends on stellar parameters such as metallicity and effective temperature, as well as on the underlying Galactic population (e.g., \citealt{2017MNRAS.465L.109F, 2019AA...624A..78D, 2020A&A...639A.127C}). Galactic chemical-evolution models have also been used to interpret the [Y/Mg]--age relation and its dependence on stellar metallicity and Galactic location \citep[e.g.,][]{2021A&A...651A..84M, 2025A&A...694A.274M}. Observational studies further show that chemical-clock relations vary across the Galactic disk, both in open clusters and in field stars with asteroseismic ages \citep[e.g.,][]{2022A&A...660A.135V, 2026A&A...713A..48M}. In particular, the tight [Y/Mg]--age relation appears to degrade outside the narrow parameter space of solar twins, and systematic offsets can arise when applying these relations to more heterogeneous samples. These limitations highlight the need for homogeneous, high-precision abundance studies of well-defined stellar populations to robustly calibrate chemical clocks and quantify their intrinsic scatter. In particular, whether and how the presence of planets affects these relations remains largely unexplored.

The Planets Around Solar Twins/Analogs (PASTA) survey \citep{2025A&A...701A.107S, 2025ApJ...980..179S} was designed to obtain high-precision elemental abundances for planet-hosting solar twins and analogs, enabling systematic studies of the interplay between planet formation, stellar evolution, and GCE. PASTA~I \citep{2025ApJ...980..179S} initiated the survey by establishing a strictly differential abundance analysis and found that the Sun remains relatively depleted in refractory elements compared to planet-hosting solar twins, while PASTA~II \citep{2025A&A...701A.107S} assessed systematic differences between instruments. In this work, we present 57 new planet-hosting targets observed with Keck/HIRES,  LBT/PEPSI, and Subaru/HDS. Combined with PASTA~I and II, the complete sample comprises 94 stars, of
which 81 are identified as planet-hosting solar twins and analogs. This sample allows us to derive chemical clock relations based on [Y/Mg], [Y/Al], [Ba/Mg], [Ba/Al], and [Sr/Mg], and to examine whether planet-hosting solar analogs follow the same chemical clock relations as nearby solar twins.
	
\section{Observations and Data Analysis}

The sample of planet-hosting solar-type stars was compiled from known exoplanet hosts and the TESS Objects of Interest (TOI; \citealt{2021ApJS..254...39G}) catalog, as described in detail in \citet{2025ApJ...980..179S}. The Keck/HIRES spectra were drawn from the TESS--Keck Survey \citep{2023AJ....166...33M} and the California Legacy Survey \citep{2021ApJS..255....8R}, comprising 18 targets with a typical resolving power of $R \gtrsim 50,000$ and signal-to-noise ratio (S/N) exceeding 100 over 3600--9000~\AA. The LBT/PEPSI spectra were obtained through queue observations in semesters 2025A and 2025B using the 300~$\mu$m fiber ($R \approx 50,000$), adding 24 targets with typical S/N $>200$ over 3830--9120~\AA. Two targets were observed with both Keck/HIRES and LBT/PEPSI, and HD~42618 was also observed with Magellan/MIKE in PASTA~II.

In addition, we observed 13 targets in semester 2026A using LBT/PEPSI and Subaru/HDS. The LBT/PEPSI spectra were obtained with the same setup as in the 2025 observing runs. The Subaru/HDS observations used the StdYb setup, with a resolving power of $R>80,000$ and typical S/N $>100$, covering 4140--6810~\AA. These observations bring the full PASTA sample to 94 stars (shown in Table \ref{tab:age}). Among these, we identify 85 solar twins and analogs, with approximately $|\Delta T_{\rm eff}| < 200$~K, $|\Delta \log g| < 0.2$~dex, and $|\Delta{\rm [Fe/H]}| < 0.2$~dex relative to the Sun. HIP~25670, HIP~44713, and HIP~54287 do not have known planets; these stars were observed in PASTA~II primarily to assess the measurement uncertainties. TOI-1581 has recently been reclassified as a nearby eclipsing binary (NEB) and is therefore no longer considered a planet-hosting target. Consequently, the final sample contains 81 planet-hosting solar twins and analogs. These 81 stars are marked with an asterisk next to their names in Table~\ref{tab:age}.

\begin{table*}[t]
	\centering
	\caption{Stellar ages, masses, and radii of the 94 stars in the PASTA sample
		\label{tab:age}}
	\resizebox{\textwidth}{!}{%
		\begin{tabular}{l|ccc|ccc|ccc|ccc|ccc|ccc}
			\hline\hline
			& \multicolumn{6}{c}{Age$^a$ (Gyr)}
			& \multicolumn{6}{c}{Mass$^a$ ($M_\odot$)}
			& \multicolumn{6}{c}{Radius$^a$ ($R_\odot$)} \\
			\cline{2-7}\cline{8-13}\cline{14-19}
			Star
			& \multicolumn{3}{c}{Plx$^a$}
			& \multicolumn{3}{c}{Spec$^a$}
			& \multicolumn{3}{c}{Plx}
			& \multicolumn{3}{c}{Spec}
			& \multicolumn{3}{c}{Plx}
			& \multicolumn{3}{c}{Spec} \\
			\cline{2-4}\cline{5-7}\cline{8-10}\cline{11-13}
			\cline{14-16}\cline{17-19}
			& MP$^b$ & 16th$^b$ & 84th$^b$
			& MP & 16th & 84th
			& MP & 16th & 84th
			& MP & 16th & 84th
			& MP & 16th & 84th
			& MP & 16th & 84th \\
			\hline
			
			TOI-1036$^*$
			& 4.500 & 3.770 & 5.033
			& 5.000 & 3.650 & 5.599
			& 1.060 & 1.047 & 1.077
			& 1.070 & 1.057 & 1.088
			& 1.080 & 1.059 & 1.110
			& 1.120 & 1.046 & 1.203 \\
			
			TOI-1055$^*$
			& 3.500 & 2.882 & 4.584
			& 7.500 & 6.041 & 8.251
			& 1.000 & 0.983 & 1.017
			& 0.990 & 0.981 & 1.004
			& 0.980 & 0.969 & 0.987
			& 1.120 & 1.038 & 1.179 \\
			
			TOI-1076
			& 1.000 & 0.432 & 2.053
			& 1.000 & 0.412 & 2.344
			& 1.090 & 1.073 & 1.109
			& 1.090 & 1.077 & 1.107
			& 1.010 & 0.991 & 1.028
			& 1.010 & 0.987 & 1.039 \\
			
			TOI-1097$^*$
			& 1.300 & 0.785 & 1.963
			& 0.600 & 0.279 & 2.374
			& 1.050 & 1.034 & 1.067
			& 1.050 & 1.038 & 1.078
			& 0.970 & 0.962 & 0.982
			& 0.980 & 0.966 & 1.020 \\
			
			TOI-1117$^*$
			& 6.100 & 5.548 & 7.005
			& 6.400 & 4.480 & 7.643
			& 1.010 & 0.993 & 1.027
			& 1.010 & 0.998 & 1.025
			& 1.060 & 1.051 & 1.073
			& 1.040 & 0.989 & 1.111 \\
			
			TOI-1203
			& 13.300 & 12.836 & 13.791
			& 12.700 & 11.572 & 13.517
			& 0.870 & 0.856 & 0.886
			& 0.880 & 0.863 & 0.916
			& 1.190 & 1.181 & 1.200
			& 1.250 & 1.146 & 1.388 \\
			
			TOI-215$^*$
			& 0.400 & 0.197 & 2.163
			& 0.600 & 0.267 & 2.573
			& 1.060 & 1.037 & 1.080
			& 1.060 & 1.039 & 1.077
			& 0.970 & 0.959 & 0.993
			& 0.970 & 0.954 & 1.008 \\
			
			TOI-2011
			& 10.200 & 9.187 & 10.929
			& 5.900 & 3.355 & 7.890
			& 0.900 & 0.883 & 0.915
			& 0.910 & 0.896 & 0.929
			& 1.020 & 1.008 & 1.036
			& 0.900 & 0.868 & 0.945 \\
			
			TOI-2426$^*$
			& 6.400 & 5.410 & 7.056
			& 8.000 & 7.086 & 8.720
			& 0.970 & 0.954 & 0.983
			& 0.970 & 0.961 & 0.991
			& 0.980 & 0.965 & 1.002
			& 1.040 & 1.009 & 1.091 \\
			
			TOI-3342
			& -- & -- & --
			& 0.500 & 0.235 & 1.276
			& -- & -- & --
			& 1.150 & 1.131 & 1.165
			& -- & -- & --
			& 1.060 & 1.048 & 1.079 \\
			
			TOI-4628$^*$
			& 0.900 & 0.448 & 1.999
			& 1.800 & 0.693 & 3.791
			& 1.020 & 1.002 & 1.038
			& 1.060 & 1.032 & 1.078
			& 0.930 & 0.917 & 0.946
			& 0.960 & 0.945 & 1.024 \\
			
			TOI-4914$^*$
			& 0.900 & 0.338 & 1.760
			& 1.700 & 0.662 & 3.515
			& 1.040 & 1.024 & 1.057
			& 1.070 & 1.044 & 1.085
			& 0.950 & 0.940 & 0.965
			& 0.980 & 0.965 & 1.042 \\
			
			TOI-5005$^*$
			& 0.300 & 0.163 & 0.716
			& 2.800 & 1.430 & 4.204
			& 1.040 & 1.025 & 1.052
			& 1.050 & 1.039 & 1.063
			& 0.920 & 0.913 & 0.933
			& 0.990 & 0.967 & 1.034 \\
			
			TOI-5795$^*$
			& 11.500 & 10.581 & 12.214
			& 11.500 & 10.279 & 12.186
			& 0.910 & 0.895 & 0.927
			& 0.920 & 0.904 & 0.936
			& 1.130 & 1.092 & 1.165
			& 1.130 & 1.046 & 1.243 \\
			
			TOI-744$^*$
			& 8.000 & 7.292 & 8.682
			& 6.600 & 4.313 & 7.927
			& 0.970 & 0.954 & 0.984
			& 0.970 & 0.960 & 0.986
			& 1.080 & 1.049 & 1.106
			& 0.980 & 0.946 & 1.039 \\
			
			... & ... & ... & ... & ... & ... & ... & ... & ... & ... & ... & ... & ... & ... & ... & ...
			& ... & ... & ... \\
			\hline
		\end{tabular}%
	}
	
	\tablefoot{
		a. Stellar ages, masses, and radii derived from Yonsei--Yale isochrone fitting using parallax (Plx) and dereddened apparent $V$ magnitude, and spectroscopic stellar parameters (Spec) including $T_{\rm eff}$, log g, and [Fe/H]. \\
		b. MP denotes the most probable estimate of age, mass, and radius, while 16th and 84th denote the 16th and 84th percentiles of the posterior distributions returned by \texttt{q$^2$}. ``Plx'' denotes the solution using the parallax and apparent $V$ magnitude, while ``Spec'' denotes the solution based on the spectroscopic stellar parameters. A dash indicates that \texttt{q$^2$} did not return a valid value for that quantity. \\
		$^{*}$. Stars marked with an asterisk are identified as planet-hosting solar twins or analogs in the PASTA sample, comprising 81 stars in total. \\
		Only the first few rows of the full table are shown here; the complete table for all 94 stars in the PASTA survey is available in electronic form at the CDS.}
\end{table*}

The stellar and planetary parameters of the 52 new targets are listed in
Tables~\ref{tab:param1}, \ref{tab:param2}, and \ref{tab:new_targets}, while those of the previous 42 targets are presented in the PASTA I and II papers. The planetary status of each target is adopted from published studies and the TESS Follow-up Observing Program (TFOP), while the planetary parameters are compiled from the literature. Stellar parameters are derived following the procedures described by \citet{2025ApJ...980..179S, 2025A&A...701A.107S}. Briefly, the atmospheric parameters are determined from the excitation and ionization balance of Fe~I and Fe~II abundances, using a strictly line-by-line differential analysis relative to the Sun. 

\begin{table*}[t]
	\centering
	\caption{Stellar and planetary parameters of 18 Keck/HIRES targets \label{tab:param1}}
	\resizebox{\textwidth}{!}{%
		\begin{tabular}{lcccccccccccc}
			\hline\hline
			Planet$^a$ & $T_{\rm eff}^b$ & $\sigma^b$ & $\log g^b$ & $\sigma^b$ &
			[Fe/H]$^b$ & $\sigma^b$ & $V_t^b$ & $\sigma^b$ & status$^c$ &
			$P_{\rm orb}^d$ & $M_p^d$ & $R_p^d$ \\
			& (K) & (K) & (dex) & (dex) &
			(dex) & (dex) & (km s$^{-1}$) & (km s$^{-1}$) & &
			(days) & ($M_{\oplus}$) & ($R_{\oplus}$) \\
			\hline
			TOI-1723 b & 5758 & 15 & 4.33 & 0.04 & 0.07 & 0.01 & 0.90 & 0.03 & KP & 13.726 & $10.4^{+5.2}_{-4.8}$ & 3.292$^{+0.143}_{-0.128}$ \\
			TOI-1799 b & 5697 & 13 & 4.38 & 0.04 & -0.01 & 0.01 & 0.87 & 0.04 & KP & 7.086 & $4.0^{+1.7}_{-1.8}$ & 1.422$^{+0.093}_{-0.088}$ \\
			TOI-1742 b & 5738 & 13 & 4.31 & 0.04 & 0.16 & 0.01 & 0.92 & 0.03 & KP & 21.269 & 9.70$\pm$1.90 & 2.365$_{-0.051}^{+0.057}$ \\
			TOI-1710 b & 5718 & 13 & 4.41 & 0.04 & 0.08 & 0.01 & 0.90 & 0.04 & KP & 24.283 & $22.4^{+4.1}_{-4.0}$ & 5.203$_{-0.091}^{+0.096}$ \\
			TOI-1691 b & 5711 & 13 & 4.40 & 0.04 & 0.05 & 0.01 & 0.83 & 0.04 & KP & 16.737 & $14.6^{+5.4}_{-5.3}$ & 3.565$_{-0.087}^{+0.107}$ \\
			TOI-1473 b & 5972 & 18 & 4.49 & 0.04 & 0.04 & 0.01 & 0.82 & 0.05 & KP & 5.255 & 10.0$\pm$2.5 & 2.428$_{-0.073}^{+0.099}$ \\
			TOI-1471 b & 5656 & 12 & 4.43 & 0.04 & 0.02 & 0.01 & 0.84 & 0.04 & KP & 20.773 & $<$8.3$\pm$1.7 & 3.828$_{-0.088}^{+0.104}$ \\
			TOI-1451 b & 5849 & 13 & 4.54 & 0.02 & 0.07 & 0.01 & 0.95 & 0.03 & KP & 16.538 & 15.2$\pm$2.8 & 2.611$_{-0.103}^{+0.134}$ \\
			TOI-1422 b & 5907 & 28 & 4.52 & 0.05 & 0.05 & 0.02 & 0.83 & 0.06 & KP & 12.999 & $9.50^{+2.00}_{-1.90}$ & 3.83$\pm$0.11 \\
			TOI-1422 c &  &  &  &  &  &  &  &  & KP & 34.563 & 14$\pm$3 & 2.61$\pm$0.14 \\
			TOI-1386 b & 5833 & 14 & 4.42 & 0.04 & 0.12 & 0.01 & 0.90 & 0.04 & KP & 25.840 & $45.8^{+5.8}_{-5.7}$ & 6.216$_{-0.170}^{+0.206}$ \\
			TOI-1386 c &  &  &  &  &  &  &  &  & KP & 232.0 & $75.2^{+15.3}_{-15.0}$* & 10.29** \\
			TOI-1247 b & 5738 & 10 & 4.49 & 0.03 & -0.15 & 0.01 & 0.91 & 0.03 & KP & 15.923 & 6.1$\pm$1.8 & 2.532$_{-0.070}^{+0.084}$ \\
			TOI-1136 b & 5879 & 29 & 4.64 & 0.05 & 0.09 & 0.02 & 1.07 & 0.07 & KP & 4.173 & $3.50^{+0.80}_{-0.70}$ & $1.90^{+0.21}_{-0.15}$ \\
			TOI-1136 c &  &  &  &  &  &  &  &  & KP & 6.257 & $7.60^{+5.10}_{-5.00}$ & 2.769$_{-0.067}^{+0.077}$ \\
			TOI-1136 d &  &  &  &  &  &  &  &  & KP & 12.519 & $8.350^{+1.800}_{-1.600}$ & 4.621$_{-0.103}^{+0.153}$ \\
			TOI-1136 e &  &  &  &  &  &  &  &  & KP & 18.807 & $6.07^{+1.09}_{-1.01}$ & 2.546$_{-0.088}^{+0.105}$ \\
			TOI-1136 f &  &  &  &  &  &  &  &  & KP & 26.318 & $9.70^{+3.90}_{-3.70}$ & 3.736$_{-0.079}^{+0.086}$ \\
			TOI-1136 g &  &  &  &  &  &  &  &  & KP & 39.539 & $12.2^{+9.3}_{-9.6}$ & $2.53^{+0.11}_{-0.12}$ \\
			HD 213519 b & 5825 & 13 & 4.38 & 0.04 & 0.03 & 0.01 & 0.91 & 0.03 & KP & 11.122 & $9.2^{+2.9}_{-1.3}$* & 2.99** \\
			HD 187123 b & 5839 & 17 & 4.40 & 0.05 & 0.14 & 0.02 & 0.99 & 0.05 & KP & 3.097 & $176^{+6}_{-10}$* & 14.23** \\
			HD 187123 c &  &  &  &  &  &  &  &  & KP & 3396$\pm$26 & $912^{+127}_{-133}$ & 13.2** \\
			HD 164595 b & 5778 & 38 & 4.55 & 0.06 & -0.04 & 0.03 & 0.92 & 0.08 & KP & 40.0 & 16.14$\pm$2.72* & 4.16** \\
			HD 42618 b & 5754 & 12 & 4.47 & 0.04 & -0.10 & 0.01 & 0.98 & 0.03 & KP & 148.49 & 15.2$\pm$ 1.8* & 3.89** \\
			HD 32963 b & 5742 & 12 & 4.33 & 0.03 & 0.08 & 0.01 & 0.90 & 0.03 & KP & 2328 & $0.726^{+0.036}_{-0.035}$* & 13.34** \\
			51 Peg b & 5807 & 13 & 4.36 & 0.03 & 0.22 & 0.01 & 0.99 & 0.03 & KP & 4.231 & $146^{+19}_{-3}$ & 14.2** \\
			\hline
		\end{tabular}%
	}
	\par\smallskip
	\footnotesize
	\noindent\textit{Notes.}
	$^a$ Name of the planet(s). Planetary parameters are adopted from the literature: TOI-1723 b, TOI-1799 b, TOI-1742 b, TOI-1710 b, TOI-1691 b, TOI-1473 b, TOI-1471 b, TOI-1451 b, TOI-1386 b, c, TOI-1247 b, and TOI-1136 b--g from \citet{2024ApJS..272...32P} (with masses for TOI-1136 from \citealt{2024AJ....167...70B}); TOI-1422 b, c from \citet{2026MNRAS.545f2030N}; HD~213519 b from \citet{2022ApJS..262...21F}; HD~187123 b, c from \citet{2025ApJS..280...61A}; HD~164595 b from \citet{2015AA...581A..38C}; HD~42618 b and HD~32963 b from \citet{2021ApJS..255....8R}; and 51~Peg~b from \citet{2015AA...576A.134M}. When radii or masses are unavailable in the discovery papers, we adopt values from the NASA Exoplanet Archive.
	
	$^b$ Stellar effective temperature ($T_{\rm eff}$), surface gravity (log $g$), microturbulence ($V_t$), metallicity ([Fe/H]), and their associated uncertainties of the host stars.
	
	$^c$ Current status of the planetary system. We adopt the TESS Follow-up Observing Program (TFOP) Working Group (WG) classifications\footnote{\url{https://tess.mit.edu/followup/}}. ``KP'' denotes a known (published) planet; ``CPC'' a disposition-cleared planetary candidate; ``VPC'' a validated planetary candidate; ``VPC$-$'' a validated planetary candidate with caveats; and ``NEB'' a nearby eclipsing binary (i.e., a false positive).
	
	$^d$ Orbital period ($P$), planet mass ($M_p$, in Earth masses $M_{\oplus}$), and planet radius ($R_p$, in Earth radii $R_{\oplus}$) are adopted from the literature as noted above. When $M_p \sin i$ is reported instead of $M_p$, these values are indicated with an asterisk (*) to the right. For systems without published values, we adopt estimates from the NASA Exoplanet Archive; these are marked with a double asterisk (**).
\end{table*}

\begin{table*}[t]
	\centering
	\caption{Stellar and planetary parameters of the 24 LBT/PEPSI targets\label{tab:param2}}
	\resizebox{\textwidth}{!}{%
		\begin{tabular}{lcccccccccccc}
			\hline\hline
			Planet$^a$ & $T_{\rm eff}$ & $\sigma$ & $\log g$ & $\sigma$ &
			[Fe/H] & $\sigma$ & $V_t$ & $\sigma$ & status &
			$P_{\rm orb}$ & $M_p$ & $R_p$ \\
			& (K) & (K) & (dex) & (dex) &
			(dex) & (dex) & (km s$^{-1}$) & (km s$^{-1}$) & &
			(days) & ($M_{\oplus}$) & ($R_{\oplus}$) \\
			\hline
			TOI-1136 b-g & 5909 & 33 & 4.64 & 0.06 & 0.083 & 0.03 & 1.05 & 0.08 & \multicolumn{4}{c}{Shown above} \\
			TOI-1247 b & 5715 & 24 & 4.34 & 0.06 & -0.052 & 0.032 & 0.61 & 0.08 & \multicolumn{4}{c}{Shown above} \\
			TOI-1860 b & 5777 & 52 & 4.60 & 0.09 & 0.028 & 0.037 & 1.19 & 0.10 & KP & 1.067 & 2.27 & 1.31$\pm$0.04 \\
			HD 9446 b & 5824 & 26 & 4.51 & 0.05 & 0.07 & 0.02 & 1.26 & 0.06 & KP & 30.052 & 222.472$\pm$19.069 & 14.0$**$ \\
			HD 9446 c &  &  &  &  &  &  &  &  & KP & 192.9 & 578.426$\pm$54.029 & 13.44$**$ \\
			HD 17674 b & 5932 & 29 & 4.29 & 0.08 & -0.15 & 0.02 & 1.04 & 0.05 & KP & 623.8 & $277^{+22}_{-19}$ & 13.9$**$ \\
			K2-185 b & 5784 & 32 & 4.47 & 0.07 & -0.12 & 0.02 & 1.20 & 0.08 & KP & 10.616 & 1.6 & 1.15$\pm$0.07 \\
			K2-185 c &  &  &  &  &  &  &  &  & KP & 52.713 & 6.3 & 2.39$\pm$0.09 \\
			K2-352 b & 5805 & 31 & 4.38 & 0.05 & 0.00 & 0.02 & 1.07 & 0.06 & KP & 3.666 & 2.45 & $1.37^{+0.10}_{-0.07}$ \\
			K2-352 c &  &  &  &  &  &  &  &  & KP & 8.235 & 4.35 & $1.92^{+0.13}_{-0.07}$ \\
			K2-352 d &  &  &  &  &  &  &  &  & KP & 14.871 & 5.6 & $2.23^{+0.13}_{-0.09}$ \\
			Kepler-96 b & 5765 & 26 & 4.46 & 0.05 & 0.03 & 0.02 & 1.15 & 0.06 & KP & 16.239 & 8.46$\pm$3.4 & 2.67$\pm$0.22 \\
			Kepler-130 b & 5905 & 48 & 4.30 & 0.12 & -0.21 & 0.03 & 1.21 & 0.09 & KP & 8.457 & 1.04 & 1.040$\pm$0.036 \\
			Kepler-130 c &  &  &  &  &  &  &  &  & KP & 27.509 & 8.75 & 2.877$\pm$0.072 \\
			Kepler-130 d &  &  &  &  &  &  &  &  & KP & 87.517 & 3.33 & 1.379$\pm$0.079 \\
			Kepler-131 b & 5824 & 27 & 4.51 & 0.06 & 0.15 & 0.02 & 1.08 & 0.07 & KP & 16.092 & 16.13$\pm$3.5 & 2.41$\pm$0.20 \\
			Kepler-131 c &  &  &  &  &  &  &  &  & KP & 25.517 & 8.25$\pm$5.9 & 0.84$\pm$0.07 \\
			TOI-963.01 & 5815 & 26 & 4.40 & 0.05 & -0.04 & 0.02 & 1.19 & 0.06 & VPC- & 11.118 & -- & 8.851$\pm$0.425 \\
			TOI-1581.01 & 5845 & 35 & 4.50 & 0.07 & -0.07 & 0.03 & 1.15 & 0.07 & NEB & 1807.492 & -- & 18.620$\pm$5.673 \\
			TOI-2010 b & 5819 & 25 & 4.40 & 0.05 & 0.13 & 0.02 & 1.03 & 0.05 & KP & 141.834 & $408.7^{+17.5}_{-18.1}$ & 11.81$\pm$0.30 \\
			TOI-4443 b & 5818 & 27 & 4.41 & 0.05 & 0.02 & 0.02 & 1.14 & 0.06 & KP & 1.850 & 3.6 & 1.719$\pm$0.135 \\
			TOI-4649.01 & 5828 & 66 & 4.50 & 0.16 & 0.05 & 0.06 & 1.20 & 0.20 & PC & 15.076 & -- & 2.419$\pm$0.205 \\
			TOI-5703.01 & 5896 & 39 & 4.43 & 0.07 & -0.09 & 0.03 & 1.24 & 0.08 & PC & 10.169 & -- & 2.485$\pm$0.163 \\
			TOI-5958.01 & 5855 & 28 & 4.48 & 0.05 & -0.02 & 0.02 & 1.19 & 0.06 & PC & 2.306 & -- & 2.065 \\
			TOI-5967.01 & 5830 & 51 & 4.52 & 0.11 & 0.03 & 0.04 & 1.18 & 0.12 & PC & 15.344 & -- & 3.325$\pm$0.203 \\
			TOI-6074.01 & 5836 & 44 & 4.52 & 0.11 & -0.03 & 0.03 & 1.06 & 0.12 & PC & 12.132 & -- & 1.569$\pm$0.398 \\
			TOI-6109 b & 5703 & 65 & 4.50 & 0.14 & -0.08 & 0.06 & 2.03 & 0.13 & KP & 5.690 & 21.1 & $4.870^{+0.162}_{-0.123}$ \\
			TOI-6109 c &  &  &  &  &  &  &  &  & KP & 8.539 & 20.8 & $4.8326^{+0.0736}_{-0.0632}$ \\
			TOI-6243.01 & 5699 & 35 & 4.44 & 0.07 & 0.20 & 0.03 & 1.19 & 0.09 & VPC & 3.565 & -- & 2.888 \\
			TOI-6250.01 & 5811 & 35 & 4.32 & 0.06 & -0.03 & 0.03 & 1.15 & 0.07 & PC & 1076.61 & -- & 2.751$\pm$3.087 \\
			TOI-6333.01 & 5875 & 34 & 4.50 & 0.08 & -0.07 & 0.02 & 1.21 & 0.08 & VPC? & 2.721 & -- & 3.022$\pm$0.225 \\
			TOI-6652.01 & 5858 & 34 & 4.53 & 0.08 & 0.07 & 0.03 & 1.30 & 0.08 & CPC & 3.839 & -- & 1.949$\pm$0.153 \\
			\hline
		\end{tabular}%
	}
	\par\smallskip
	\footnotesize
	\noindent\textit{Notes.}
	$^a$ Name of the planet(s). Planetary parameters are adopted from the literature: TOI-1860 b from \citet{2022AJ....163...99G}; HD~9446 b, c from \citet{2010AA...513A..69H}; HD~17674 b from \citet{2017AA...601A...9R}; K2-185 b, c and K2-352 b, c, d from \citet{2021MNRAS.508..195D}; Kepler-96 b and Kepler-131 b, c from \citet{2014ApJS..210...20M}; Kepler-130 b, c, d from \citet{2024ApJS..270....8W}; TOI-2010 b from \citet{2023AJ....166..239M}; TOI-4443 b from \citet{2024AJ....167..233H}; and TOI-6109 b, c from \citet{2025AJ....170..318D}. For TOI candidates without confirmed planetary parameters, values are adopted from ExoFOP. When radii or masses are unavailable in the discovery papers, we adopt values from the NASA Exoplanet Archive. Values of $M_p\sin i$ are indicated with an asterisk (*), while parameters adopted from the NASA Exoplanet Archive are marked with a double asterisk (**). The rest of the columns are the same as Table \ref{tab:param1}.
\end{table*}

\begin{table*}
	\caption{Stellar and planetary parameters of the 13 targets observed in semester 2026A \label{tab:new_targets}}
	\footnotesize
	\centering
	\begin{tabular}{lcccccccccccc}
		\hline\hline
		Target$^{a}$ & $T_{\rm eff}$ & $\sigma$ & $\log g$ & $\sigma$ & [Fe/H] & $\sigma$ & $V_t$ & $\sigma$ & status & $P_{\rm orb}$ & $M_p$ & $R_p$ \\
		& (K) & (K) & (dex) & (dex) & (dex) & (dex) & (km s$^{-1}$) & (km s$^{-1}$) & & (days) & ($M_{\oplus}$) & ($R_{\oplus}$) \\
		\hline
		\multicolumn{13}{c}{LBT/PEPSI} \\
		\hline
		K2-418 b$^{b}$ & 5865 & 25 & 4.48 & 0.07 & $-0.137$ & 0.02 & 0.97 & 0.07 & KP & $16.141$ & $10.4^{+1.6}_{-1.5}$ & $2.332^{+0.080}_{-0.094}$ \\
		HD 106252 b$^{b}$ & 5928 & 23 & 4.45 & 0.07 & $-0.089$ & 0.02 & 1.04 & 0.06 & KP & 1535 & $3178^{+248}_{-232}$ & $12.47$ \\
		HD 109286 b$^{b}$ & 5770 & 32 & 4.65 & 0.06 & $+0.114$ & 0.03 & 0.78 & 0.09 & KP & 520.1 & $950^{+48}_{-48}$ (sin i) & $13.14$ \\
		TOI-1687.01$^{b}$ & 5879 & 83 & 4.70 & 0.17 & $+0.072$ & 0.07 & 1.49 & 0.16 & PC & $10.26$ & $18.03^{+6.77}_{-5.53}$ & $3.913\pm0.289$ \\
		TOI-1772 b$^{b}$  & 5715 & 35 & 4.54 & 0.06 & $+0.070$ & 0.03 & 0.75 & 0.09 & KP & $8.054$ & $8.78$ & $2.905^{+0.328}_{-0.214}$ \\
		TOI-1772.02$^{b}$ &      &    &      &      &          &      &      &      & PC & 4.726 & $0.1$ & $0.422\pm0.035$ \\
		TOI-1777.01$^{b}$ & 5821 & 42 & 4.46 & 0.12 & $-0.044$ & 0.04 & 0.95 & 0.11 & PC & $14.65$ & $5.67$ & $2.246^{+0.279}_{-0.177}$ \\
		TOI-2091.01$^{b}$ & 5887 & 54 & 4.52 & 0.09 & $+0.084$ & 0.04 & 0.99 & 0.11 & PC & 177.219 & $5.74^{+6.36}_{-4.14}$ & $2.032\pm1.132$ \\
		TOI-2289.01$^{b}$ & 5696 & 19 & 4.39 & 0.06 & $-0.056$ & 0.02 & 0.66 & 0.07 & PC & 63.395 & $12.54^{+5.46}_{-4.04}$ & $3.351\pm0.394$ \\
		TOI-5726.01$^{b}$ & 5878 & 37 & 4.39 & 0.10 & $+0.181$ & 0.03 & 0.94 & 0.09 & PC & $5.491$ & $7.05$ & $2.554^{+0.280}_{-0.177}$ \\
		\hline
		\multicolumn{13}{c}{Subaru/HDS} \\
		\hline
		TOI-2070.01$^{b}$ & 5935 & 53  & 4.54 & 0.09 & $+0.11$ & 0.04 & 1.15 & 0.10 & PC & $6.384$ & $8.24^{+15.46}_{-6.24}$ & $2.284\pm2.195$ \\
		TOI-2073.01$^{b}$ & 5802 & 46  & 4.35 & 0.12 & $+0.14$ & 0.04 & 1.20 & 0.09 & PC & $53.199$ & $8.89^{+10.41}_{-5.19}$ & $2.84\pm1.169$ \\
		TOI-1734.01$^{b}$ & 5852 & 47  & 4.35 & 0.13 & $+0.14$ & 0.04 & 1.00 & 0.11 & PC & $28.875$ & $8.11^{+2.49}_{-2.01}$ & $2.712\pm0.196$ \\
		KELT-23 A b$^{b}$ & 5865 & 53  & 4.36 & 0.13 & $+0.07$ & 0.04 & 0.93 & 0.12 & KP & $2.255$ & $298^{+15}_{-14}$ & $14.83\pm0.28$ \\
		\hline
	\end{tabular}
	\tablefoot{
		$^{a}$ Planet name for targets with a known planet or planet candidate; the stellar identifier is given for targets without a listed planet. ``KP'' denotes a known planet, while ``PC'' denotes a planet candidate. \\
		$^{b}$ Planetary parameters are adopted from \citet{2023AA...677A..33B} for K2-418 b; \citet{2021AJ....162..266L} for HD~106252 b; \citet{2021AA...653A..41D} for HD~109286 b; \citet{2025ApJ...994..184C} for TOI-1772 b, TOI-1777.01, and TOI-5726.01; \citet{2019AJ....158...78J} for KELT-23 A b; and ExoFOP for TOI-1687.01, TOI-1772.02, TOI-2091.01, TOI-2289.01, TOI-2070.01, TOI-2073.01, and TOI-1734.01. Values without quoted uncertainties are reported as provided in the source catalog. \\
	}
\end{table*}

\begin{figure}[!htbp]
	\centering
	\includegraphics[width=0.5\textwidth]{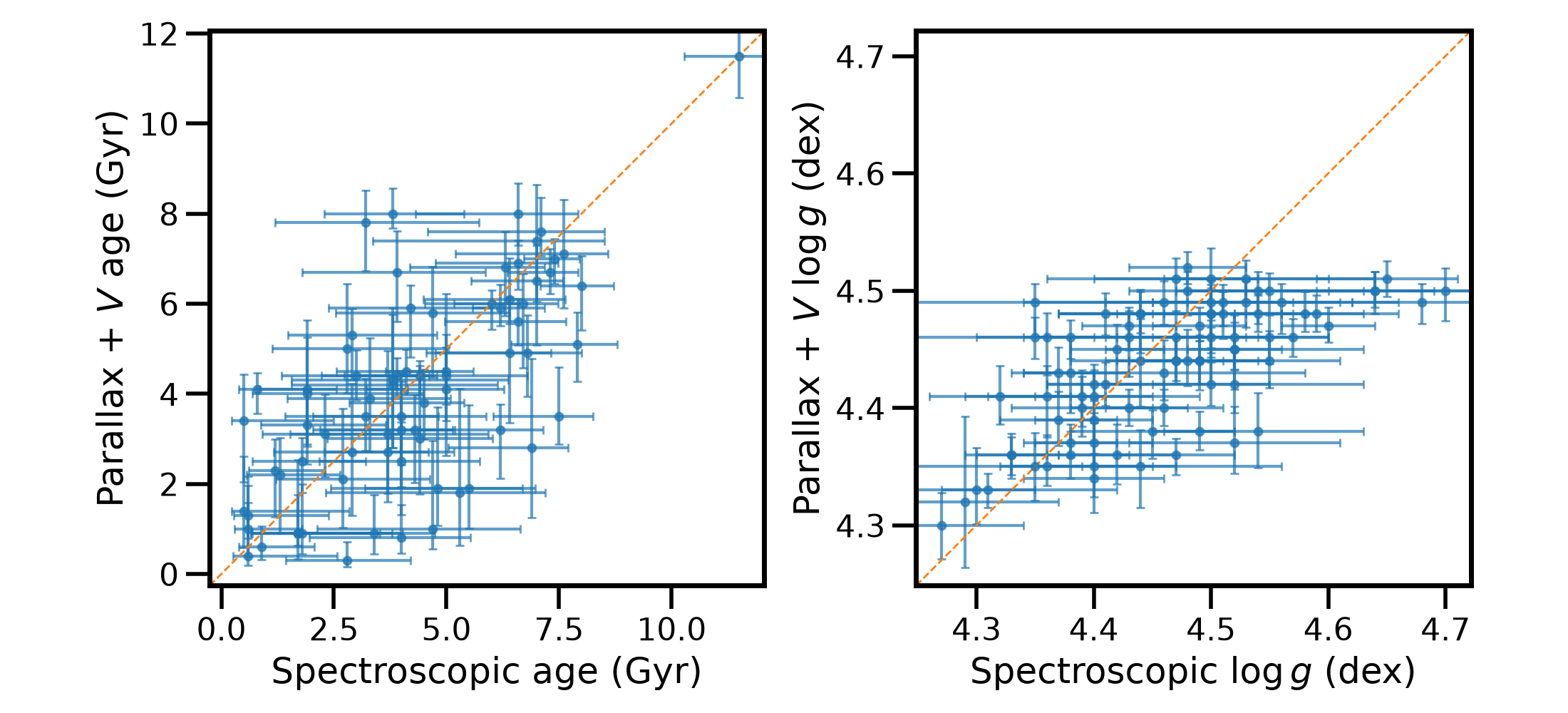}
	\caption{Comparison of parallax-based and spectroscopic ages (left) and $\log g$ values (right) for the PASTA solar twins and analogs. The dashed lines indicate one-to-one agreement.}
	\label{fig:comp}
\end{figure}

Stellar ages, masses, and radii are then derived by fitting Yonsei--Yale isochrones \citep{2001ApJS..136..417Y} with the \texttt{q$^2$} package \citep{Ramirez2014}, using $T_{\rm eff}$, $\log g$, and [Fe/H] together with their uncertainties as inputs. The fitting compares the location of each star in stellar parameter space with a given set of isochrones. The typical age uncertainties are 0.5--2.0~Gyr. For comparison and as an independent check, we also derived stellar ages using the dereddened $V$ magnitude and parallax using \texttt{q$^2$}. The most probable ages (\texttt{age\_mp}), together with the 16th and 84th percentiles, derived for the 94 targets in the PASTA sample using both the parallax-based and spectroscopic methods are presented in Table~\ref{tab:age}. We note that in PASTA I and II, we adopted the mean ages returned by the \texttt{q$^2$} package. Here, we instead adopt the most probable ages, which provide a more appropriate representation of the age posterior distributions, and use these values throughout the analysis. In Figure~\ref{fig:comp}, we compare the parallax-based ages and $\log g$ values with those derived using spectroscopic parameters alone for the 85 solar twins and analogs in our sample. The two methods show good overall agreement, providing an independent check on our age determinations. We adopt the ages derived using the spectroscopic $\log g$ throughout the subsequent analysis and discussion.

Since our sample consists of unevolved solar-type main-sequence stars, isochrone ages for individual stars are inherently uncertain because of the degeneracy of stellar evolutionary tracks near the main sequence. The overlapping targets TOI-1136 and HD~42618 yield consistent ages ($<$ 1$\sigma$) within the uncertainties, whereas TOI-1247 shows a larger discrepancy between independent observations. Although individual ages remain uncertain, they remain suitable for investigating statistical chemical clock relations across the sample. We inspected the available TESS light curves for all 94 stars but found no detectable asteroseismic oscillations that could provide independent age constraints. For overlapping targets, we adopt the Keck/HIRES results throughout this work.

\begin{figure}[!htbp]
	\centering
	\includegraphics[width=0.5\textwidth]{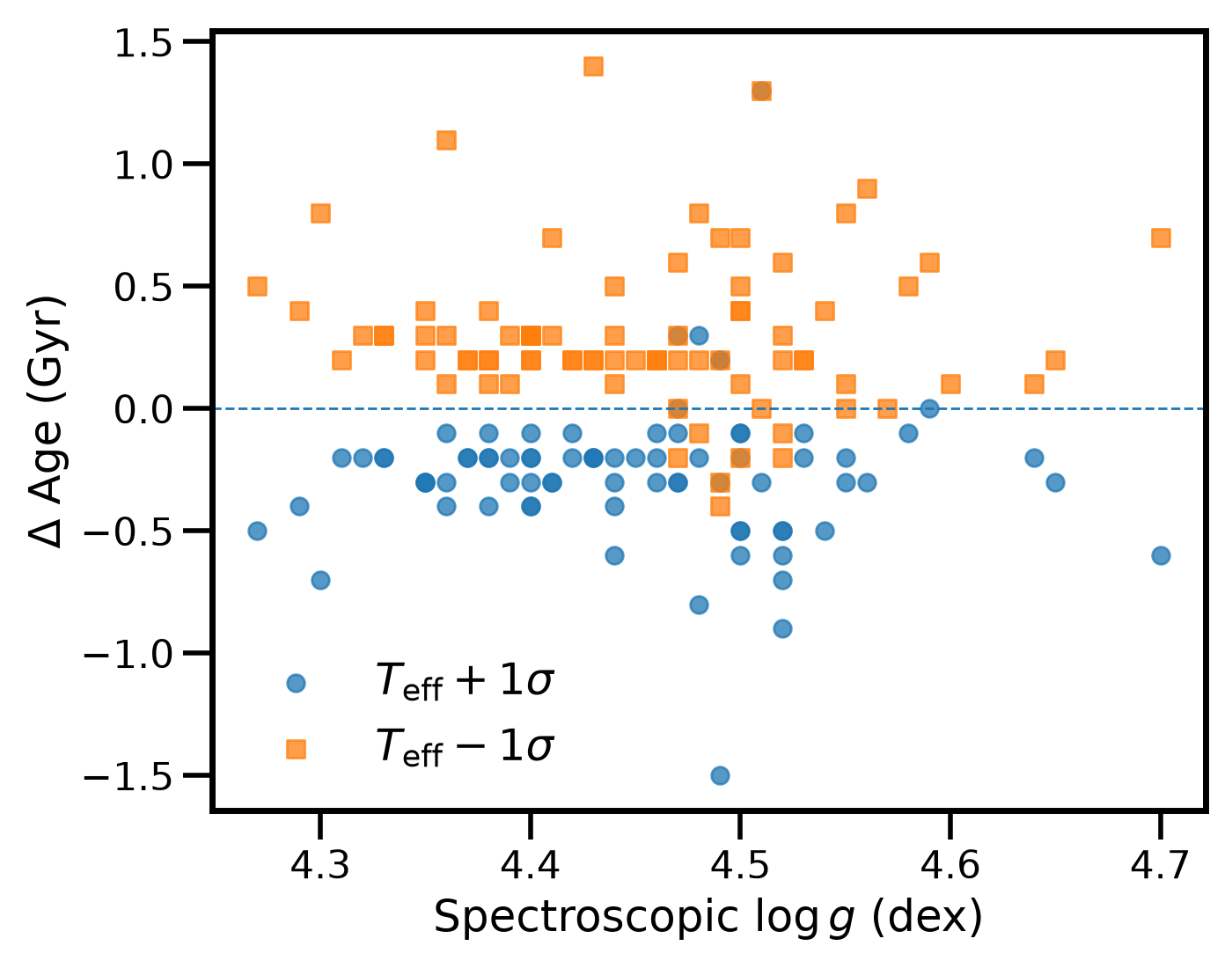}
	\caption{Changes in the spectroscopic-$\log g$-based stellar ages after varying $T_{\rm eff}$ by $+1\sigma$ (circles) and $-1\sigma$ (squares), while keeping all other input parameters unchanged, as a function of spectroscopic $\log g$ for the 85 solar twins and analogs. The age differences are calculated relative to the nominal ages. The dashed horizontal line indicates no change in the derived age.}
	\label{fig:teff_age}
\end{figure}

To assess the sensitivity of the ages derived using the spectroscopic $\log g$ to the input parameters, we repeated the isochrone fitting after varying $T_{\rm eff}$ by $+1\sigma$ and $-1\sigma$, while keeping all other input parameters unchanged. Figure~\ref{fig:teff_age} shows the resulting age changes relative to the nominal spectroscopic-$\log g$-based ages as a function of spectroscopic $\log g$ for the 85 solar twins and analogs. As expected, increasing $T_{\rm eff}$ generally results in younger inferred ages, while decreasing $T_{\rm eff}$ results in older ages. For most stars, the resulting age changes are within approximately 0.5~Gyr, smaller than the typical 16th--84th percentile intervals of the derived ages, although a few stars show differences of up to $\sim$1--1.5~Gyr. We find no systematic increase in age sensitivity toward higher $\log g$.

For the youngest systems, external age constraints are adopted when available. TOI-6109 is assigned an age of $75\pm5$ Myr based on its membership in the Alpha Persei Cluster \citep{2025AJ....170..318D}. HD~70573 (reported in PASTA~II) has been associated with several young moving groups and exhibits age estimates between $\sim20$ and 200 Myr based on kinematics and stellar activity \citep{2001MNRAS.328...45M, 2006ApJ...643.1160L, 2007ApJ...660L.145S}; we adopt an age of 100$\pm$50 Myr. No other targets in our sample are identified as members of known clusters or moving groups. These external constraints are adopted only for the youngest systems, where isochrone ages are particularly uncertain, and therefore have negligible impact on the overall chemical clock calibration.

Elemental abundances were determined following the procedures described by \citet{2025ApJ...980..179S, 2025A&A...701A.107S}. Briefly, abundances for elements with $Z\le30$ were derived from equivalent-width (EW) measurements using MARCS model atmospheres \citep{2008AA...486..951G} and the {\it abfind} driver in MOOG, while the neutron-capture elements Sr, Y, Ba, and Eu were determined by spectrum synthesis using the {\it synth} driver. All elemental abundances were derived through a strictly differential, line-by-line analysis relative to the Sun. Abundance uncertainties were computed by combining, in quadrature, the contributions from stellar atmospheric parameters and measurement uncertainties following the same procedures as in the previous PASTA papers (see also \citealt{2022MNRAS.513.5387S, 2023ApJ...952...71S, 2024AJ....167..167S, 2025ApJ...978..107S, 2025NatCo..16.9729S, 2026ApJS..282...37S, 2026ApJS..284...57S}). The measured EWs and final abundances are presented in Tables~\ref{tab:EW} and \ref{tab:abund}.

\section{Results}

\begin{figure*}[!htbp]
	\centering
	\includegraphics[width=\textwidth]{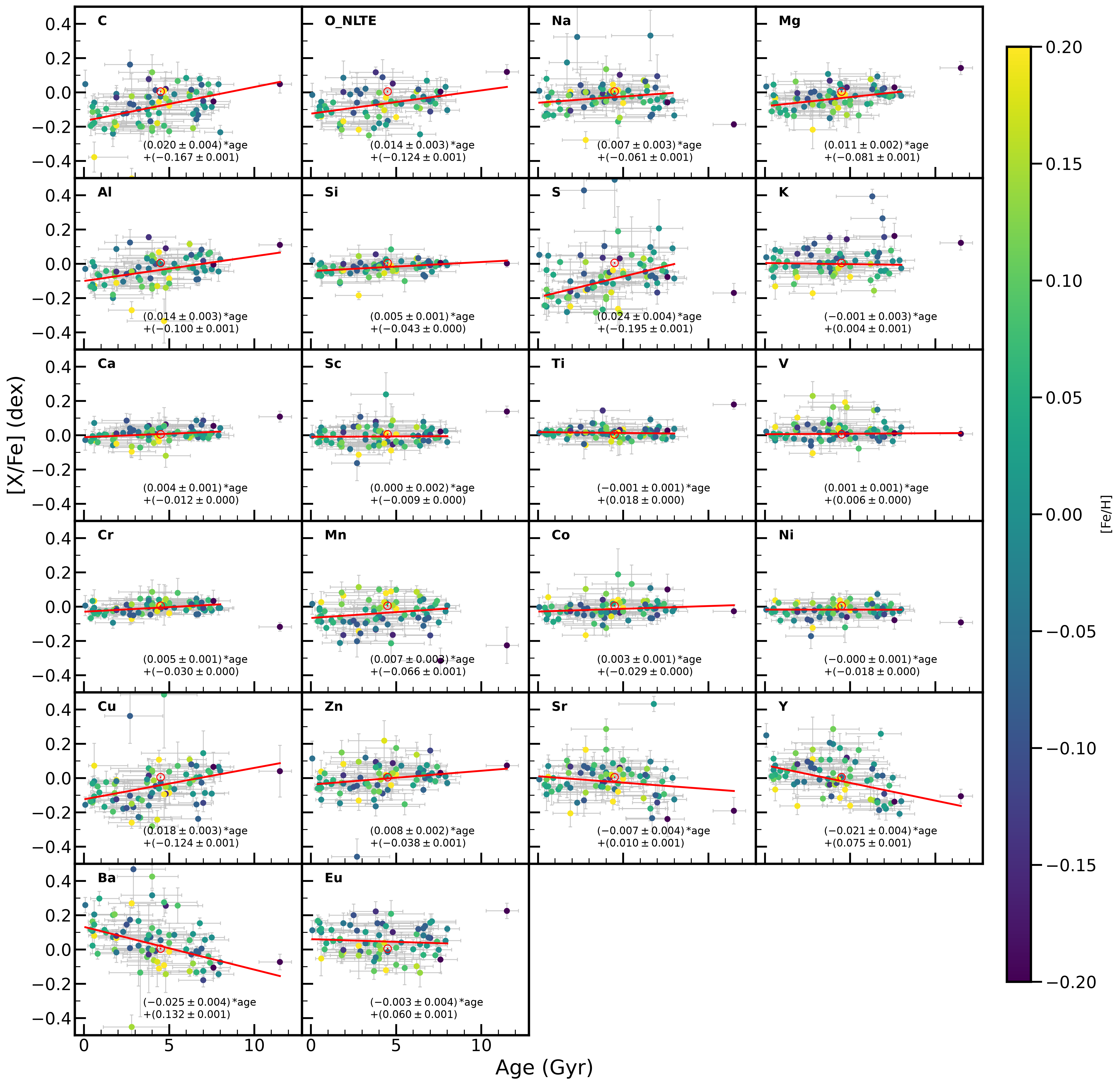}
	\caption{GCE relations for the 81 planet-hosting solar twins/analogs, showing [X/Fe] as a function of stellar age for 22 elements. Data points are color-coded by metallicity ([Fe/H]), and error bars are shown in both coordinates. The red lines represent the best-fit linear relations after $2\sigma$ clipping, with the fitted parameters listed in each panel. The solar reference point at an age of 4.5~Gyr and [X/Fe] = 0.0 dex is shown in each subplot.}
	\label{fig:synthesis}
\end{figure*}

Figure~\ref{fig:synthesis} presents the abundance patterns as a function of stellar age for the 81 planet-hosting solar analogs. Elements predominantly produced by massive stars and released on short timescales, such as Mg and S, show positive correlations with age, whereas the $s$-process elements Sr, Y and Ba show negative correlations. Similar positive slopes are observed for C, O, and Al, albeit with larger scatter, while the Fe-peak elements and the $r$-process element Eu exhibit little or no significant age dependence. These trends are consistent with the expected enrichment histories of the corresponding nucleosynthetic channels (e.g., \citealt{1979ApJ...229.1046T, 1986A&A...154..279M, 1999ARA&A..37..239B, 2014PASA...31...30K, 2006ApJ...653.1145K}) and with previous high-precision studies of nearby solar twins \citep[e.g.,][]{2012A&A...542A..84D, 2016A&A...585A.152S, 2018ApJ...865...68B}.

The opposite age dependence of the $\alpha$- and $s$-process elements reflects their different nucleosynthetic timescales. While $\alpha$-elements are predominantly produced by massive stars and enrich the interstellar medium on short timescales, the delayed enrichment from asymptotic giant branch (AGB) stars progressively increases the abundances of the $s$-process elements toward younger stellar populations \citep{1999ARA&A..37..239B,2014PASA...31...30K}. In contrast, the $r$-process element Eu exhibits little or no significant correlation with age over the parameter range covered by our sample. Similarly, the Fe-peak elements (e.g., Cr, Ni, Co, and V) show weak or negligible age dependence, as they are synthesized in both CCSNe and Type Ia supernovae and therefore closely track Fe throughout the chemical evolution of the thin disk \citep{2006ApJ...653.1145K, 2014A&A...562A..71B}. The contrasting GCE trends of the $\alpha$- and $s$-process elements naturally motivate the abundance ratios adopted as chemical clocks.

\begin{figure*}[!htbp]
	\centering
	\includegraphics[width=1.0\textwidth]{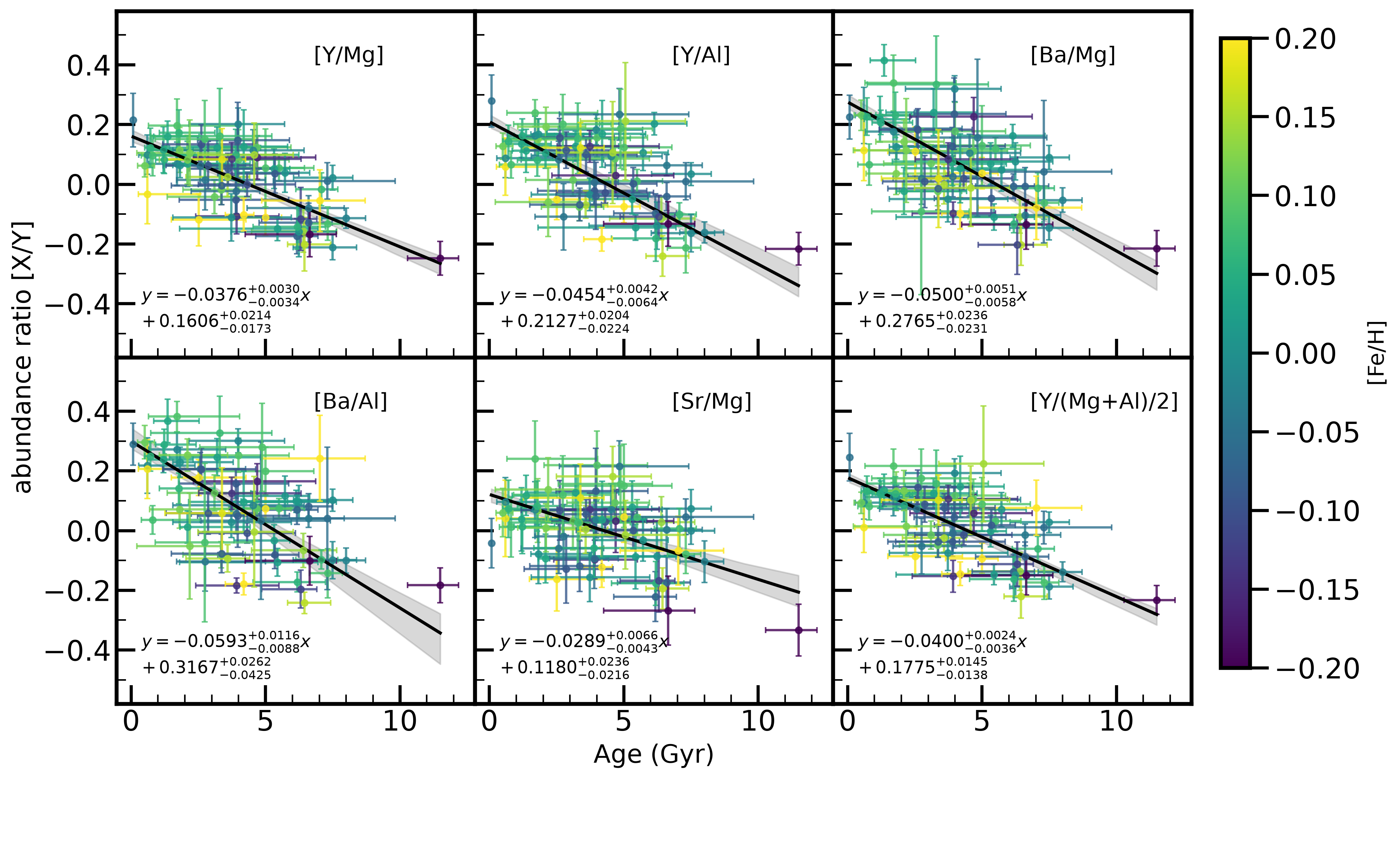}
	\caption{Chemical clock relations for planet-hosting solar analogs, showing abundance ratios as a function of stellar age for [Y/Mg], [Y/Al], [Ba/Mg], [Ba/Al], [Sr/Mg], and [Y/$(\mathrm{Mg}+\mathrm{Al})/2$]. Data points are color-coded by metallicity ([Fe/H]). Black lines show best-fit linear relations derived using Deming regression, accounting for uncertainties in both variables, and shaded regions indicate 1$\sigma$ confidence intervals. All abundance ratios show statistically significant correlations with stellar age.}
	\label{fig:clock}
\end{figure*}

Figure~\ref{fig:clock} presents the [Y/Mg], [Y/Al], [Ba/Mg], [Ba/Al], [Sr/Mg], and [Y/(Mg+Al)/2] chemical clock relations for the 81 planet-hosting solar twins/analogs in the PASTA sample. All abundance ratios exhibit clear linear correlations with stellar age, with [Y/Mg] and [Y/Al] providing the tightest relations. The fitted slopes are statistically significant relative to the null hypothesis of no correlation. We therefore adopt these abundance ratios as empirical chemical clocks.

Because stellar age uncertainties are non-negligible and comparable to the
intrinsic scatter in some abundance ratios, we adopt a Deming regression
\footnote{The regression assumes an underlying linear relation $y=a+bx$ and simultaneously infers the slope, intercept, and intrinsic scatter while accounting for measurement uncertainties in both variables. The asymmetric age uncertainties, defined by the 16th and 84th percentiles, are treated separately in the regression.} that accounts for uncertainties in both the age and abundance directions. In contrast, most previous chemical clock studies consider no or only y-axis uncertainties, which can bias the inferred slopes when age uncertainties are significant (so-called attenuation bias; \citealt{2025OJAp....8E..95T,2026OJAp....959559T}). Deming regression mitigates this effect by treating both variables symmetrically. The resulting chemical clock relations are shown in Figure~\ref{fig:clock}. 

\begin{table}
	 \centering 
	 \caption{Comparison of chemical clock relations \label{tab:slope}} \small \begin{tabular}{ccccp{3cm}} 
	 	\hline $a'$ & $b'$ & $\Delta a'$ & $\Delta b'$ & Source \\ 
	 	\hline \multicolumn{5}{c}{$\mathrm{[Y/Mg]}$} \\ \hline 0.161 & $-0.0376$ & 0.019 & 0.0032 & this work, Deming regression \\
        0.163 & $-0.0376$ & 0.019 & 0.0047 & Deming regression, excluding TOI-5795 \\ 
	 	0.146 & $-0.0332$ & 0.021 & 0.0043 & this work, considering y-err only \\
	 	0.145 & $-0.0327$ & 0.022 & 0.0047 & considering y-err only, excluding TOI-5795 \\ 
	 	0.175 & $-0.0404$ & 0.011 & 0.0019 & \citet{2015AA...579A..52N} \\ 0.170 & $-0.0371$ & 0.009 & 0.0013 & \citet{2016AA...593A..65N} \\ 0.150 & $-0.0347$ & 0.007 & 0.0012 & \citet{2017AA...608A.112N} \\ 0.176 & $-0.0410$ & 0.011 & 0.0017 & \citet{2016AA...593A.125S} \\ 0.186 & $-0.0410$ & 0.008 & 0.0010 & \citet{2016AA...590A..32T} \\ 0.209 & $-0.041$ & 0.023 & 0.003 & \citet{2019AA...624A..78D} \\ 0.186 & $-0.0410$ & 0.008 & 0.0010 & \citet{2022ApJ...936..100B} \\ 
	 	\hline \multicolumn{5}{c}{$\mathrm{[Y/Al]}$} \\ \hline 0.213 & $-0.0454$ & 0.021 & 0.0053 & this work, Deming regression \\ 
	 	0.221 & $-0.0493$ & 0.022 & 0.0073 & Deming regression, excluding TOI-5795 \\ 
	 	0.173 & $-0.0343$ & 0.026 & 0.0055 & this work, considering y-err only \\
	 	0.174 & $-0.0348$ & 0.027 & 0.0059 & considering y-err only, excluding TOI-5795 \\
	 	0.196 & $-0.0427$ & 0.009 & 0.0014 & \citet{2016AA...593A..65N} \\ 0.174 & $-0.0400$ & 0.008 & 0.0012 & \citet{2017AA...608A.112N} \\ 0.194 & $-0.0459$ & 0.011 & 0.0018 & \citet{2016AA...593A.125S} \\ 0.210 & $-0.042$ & 0.024 & 0.004 & \citet{2019AA...624A..78D} \\ 
	 	\hline 
	 	\end{tabular} \vspace{2mm} \parbox{\linewidth}{\scriptsize Coefficients $a'$ and $b'$ correspond to linear relations of the form $[\mathrm{Y}/\mathrm{X}] = a' + b' \times \mathrm{Age}$, where Age is in Gyr. The quoted uncertainties $\Delta a'$ and $\Delta b'$ are taken from the respective studies. Differences among studies primarily reflect sample selection and analysis methods. } 
 	\end{table}

\section{Discussion}

Table~\ref{tab:slope} compares the chemical clock relations derived in this work with previous calibrations from the literature. Besides the Deming regression adopted throughout this work, we also perform linear regressions considering only y-axis uncertainties to facilitate comparison with previous studies. Neglecting age uncertainties systematically produces shallower slopes, whereas the regressions considering only y-axis uncertainties remain consistent within $\sim1\sigma$ with previous calibrations based on nearby solar twins and larger stellar samples \citep[e.g.,][]{2015AA...579A..52N, 2016AA...593A.125S, 2019AA...624A..78D, 2022ApJ...936..100B}. Among these, our [Y/Mg] relation is closest to the asteroseismic calibration of \citet{2017AA...608A.112N}. The consistency between our sample, which consists exclusively of planet-hosting solar analogs, and previous calibrations suggests that planet-hosting solar analogs follow the same chemical clock relations as nearby solar twins. This does not exclude subtle abundance differences associated with planet formation, but indicates that any such effects are smaller than the current observational uncertainties.

Independent age constraints from asteroseismology or open clusters provide valuable benchmarks for chemical clock calibration \citep{2018MNRAS.475.5487S,2020A&A...639A.127C}, but such constraints are unavailable for most stars in our sample. In addition, radial migration introduces intrinsic scatter because stars observed in the solar neighborhood were not necessarily born at the same Galactocentric radius \citep[e.g.,][]{2002MNRAS.336..785S, 2009MNRAS.396..203S, 2019ApJ...884...99F}. Our sample is restricted to solar analogs within a narrow metallicity range ($-0.2\lesssim\mathrm{[Fe/H]}\lesssim+0.2$), which minimizes these effects but does not eliminate them.

Chemical-clock relations may depend on the Galactic population. For example, \citet{2021A&A...649A.126T} found different [Y/Mg]--age relations for the thin and thick disks. To assess the Galactic population membership of our sample, we perform a kinematic classification for the 81 planet-hosting solar twins and analogs. We retrieve the equatorial coordinates, parallaxes, proper motions, and radial velocities ($\alpha$, $\delta$, $\varpi$, $\mu_{\alpha *}$, $\mu_{\delta}$, and RV) from Gaia DR3 \citep{2023A&A...674A...1G}. We use these quantities to calculate the heliocentric Galactic space velocities $(U,V,W)$, adopting the convention that $U$ is positive toward the Galactic center, $V$ in the direction of Galactic rotation, and $W$ toward the north Galactic pole.

\begin{figure*}[!htbp]
	\centering
	\includegraphics[width=\textwidth]{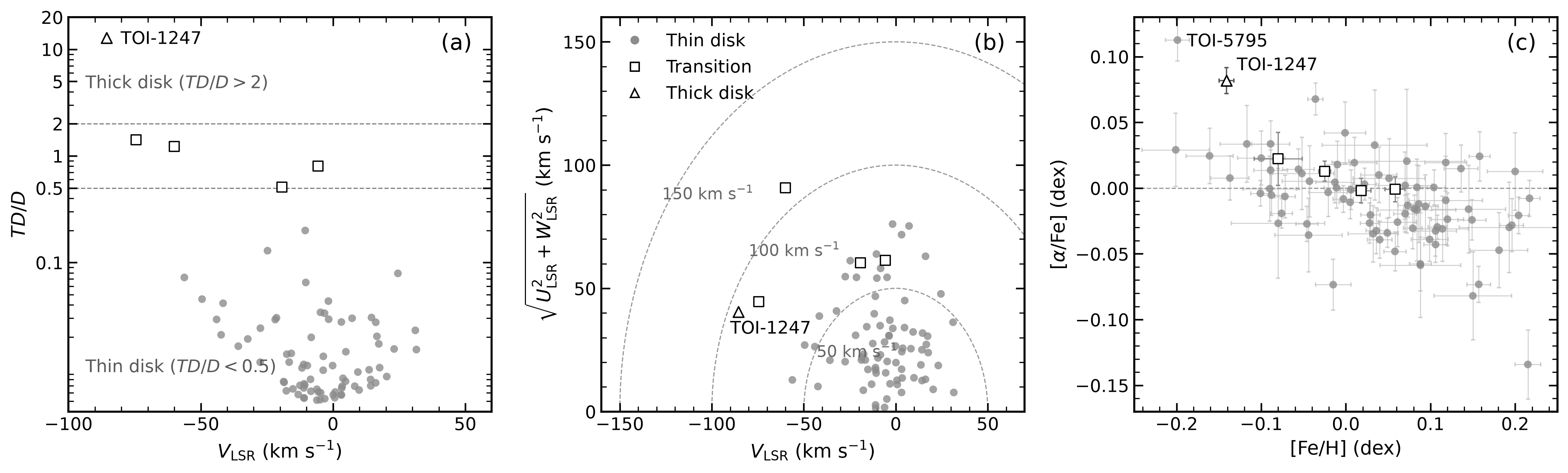}
	\caption{Kinematic and chemical properties for the 81 planet-hosting solar
		twins and analogs in PASTA. Panel (a) shows the thick-to-thin disk probability ratio, $TD/D$, as a function of $V_{\rm LSR}$, with the dashed horizontal lines at $TD/D=0.5$ and 2 separating the thin-disk, transition, and thick-disk populations. Panel (b) shows the Toomre diagram, $\sqrt{U_{\rm LSR}^{2}+W_{\rm LSR}^{2}}$ versus $V_{\rm LSR}$; the dashed curves indicate constant total LSR velocities of 50, 100, and 150 km~s$^{-1}$. Panel (c) shows [$\alpha$/Fe] as a function of [Fe/H], where the mean $\alpha$ abundance is calculated from Mg, Si, Ca, and Ti in linear abundance space. Circles, squares, and triangles denote thin-disk, transition, and thick-disk stars, respectively, based on their $TD/D$ probabilities. TOI-1247, the only thick-disk candidate in the sample, is labeled in all three panels.}
		\label{fig:kine}
	\end{figure*}

The velocities are corrected to the local standard of rest (LSR) by adopting the solar motion $(U_{\odot},V_{\odot},W_{\odot}) = (11.10,12.24,7.25)$ km~s$^{-1}$ from \citet{2010MNRAS.403.1829S}. We then calculate the relative probabilities of belonging to the thin disk, thick disk, and halo following the kinematic prescription of \citet{2014A&A...562A..71B}. This method assumes Gaussian velocity distributions for the different Galactic populations, taking into account their velocity dispersions, asymmetric drifts, and local population fractions. We use the resulting thick-to-thin disk probability ratio, $TD/D$, to classify the stars as thin-disk ($TD/D<0.5$), transition ($0.5\leq TD/D\leq2$), or thick-disk ($TD/D>2$) objects. Figure~\ref{fig:kine} shows the $TD/D$ probability ratios and Toomre diagram for the 81 planet-hosting solar twins and analogs. We find TOI-1247 ($TD/D>10$) to be a high-probability thick-disk star, while the remaining stars are likely local thin-disk stars or lie in the transition region.

We also examine the distribution of the stars in the [$\alpha$/Fe]--[Fe/H] plane. We define [$\alpha$/Fe] as the mean abundance from Mg, Si, Ca, and Ti, calculated in the $N_X$ space, and plot [$\alpha$/Fe] against [Fe/H] in Figure \ref{fig:kine}. TOI-1247 also appears relatively metal-poor and $\alpha$-enhanced compared with the bulk of the sample, consistent with its kinematic classification. TOI-5795 shows a relatively high mean [$\alpha$/Fe], mainly driven by enhanced Mg, Ca, and Ti, while [Si/Fe] remains close to solar; its kinematics ($TD/D = 0.042$), however, are consistent with the thin disk.

We repeat all chemical-clock fits after excluding TOI-1247 and find that the resulting slopes and intercepts remain essentially unchanged from those obtained with the full sample. This is also evident in Figure~\ref{fig:clock}, where TOI-1247 follows the general abundance--age trends. In contrast, TOI-5795, with an age $>10$~Gyr, is the most prominent outlier in several of the chemical-clock relations, although its kinematics are consistent with the thin disk population. We therefore also repeat the fits excluding TOI-5795. The [Y/Mg] relation remains nearly unchanged (Table \ref{tab:slope}), while the other relations show only minor differences, with all fitted slopes and intercepts remaining consistent within $1\sigma$.

Our sample is spatially confined to the solar neighborhood, spanning heliocentric distances of 15.5--317.0~pc and Galactocentric radii of $R_{\rm GC}=8.02$--8.40~kpc. We therefore use ``local'' to refer to this nearby region around the Sun. Given this relatively narrow range in $R_{\rm GC}$, the chemical-clock relations derived here primarily apply to local solar twins and analogs and should not be assumed to represent the entire Galactic disk, as spatial variations in chemical-clock relations have been found in both open clusters and field stars \citep{2022A&A...660A.135V, 2026A&A...713A..48M}.

Among the chemical clocks considered here, [Ba/Al] shows the steepest age dependence, with a slope of $-0.059$~dex~Gyr$^{-1}$. This is consistent with \citet{2022A&A...660A.135V}, who found [Ba/Al] to be among the best-performing chemical clocks in their independent sample of open clusters, showing high correlation with age and good accuracy and precision in recovering cluster ages. The agreement between these two different samples further supports [Ba/Al] as a particularly sensitive chemical clock. More generally, our steeper slopes for the Ba-based ratios compared with their Y-based counterparts are also consistent with their finding that [Ba/$\alpha$] ratios are more sensitive to age than [Y/$\alpha$] ratios.
	
\section{Conclusions}

We present high-precision stellar parameters and elemental abundances for
57 new planet-hosting targets observed with Keck/HIRES, LBT/PEPSI, and
Subaru/HDS. Combined with the samples from PASTA I and II, the present study includes 94 stars, of which 85 are solar twins/analogs and 81 are planet-hosting solar twins/analogs.

The abundance patterns follow the expected Galactic chemical evolution trends, with $\alpha$-elements exhibiting positive correlations with stellar age, whereas the $s$-process elements Y, Ba, and Sr show negative correlations. In contrast, the Fe-peak elements and the $r$-process element Eu display weak or negligible age dependence.

We derive chemical clock relations based on [Y/Mg], [Y/Al], [Ba/Mg], [Ba/Al], [Sr/Mg], and [Y/$(\mathrm{Mg}+\mathrm{Al})/2$]. These relations are consistent with previous chemical clock calibrations based on nearby solar twins, indicating that planet-hosting solar analogs follow the same chemical clock relations. Although subtle abundance signatures associated with planet formation cannot be excluded, any such effects are smaller than the current observational uncertainties.

The planet-hosting solar twins and analogs in PASTA are confined to the solar neighborhood ($d=15.5$--317.0~pc and $R_{\rm GC} = 8.02$--8.40~kpc), with the majority belonging to the Galactic thin disk and TOI-1247 being the only high-probability thick-disk star. Including or excluding TOI-1247 does not change the derived chemical-clock relations. Among the chemical clocks investigated here, [Ba/Al] shows the steepest age dependence, while the Ba-based ratios are generally more sensitive to age than their Y-based counterparts. The relations derived here should therefore be regarded primarily as calibrations for local solar twins and analogs rather than for the Galactic disk as a whole.
	
\section*{acknowledgements}
	
We thank an anonymous referee for the constructive comments and suggestions that helped improve the manuscript. This work is supported by the National Key R\&D Program of China under Grant No. 2024YFA1611801, the Science and Technology Commission of Shanghai Municipality under Grant No. 25ZR1402244, and the Shanghai Jiao Tong University Funds Program No. AF4260012. This work is also supported in part by Office of Science and Technology, Shanghai Municipal Government (grant Nos. 24DX1400100, ZJ2023-ZD-001). We acknowledge the graduate course ASTR6006H “Radiative Processes in Astrophysics” at Shanghai Jiao Tong University for its support of this work.
	
Some of the data presented herein were obtained at Keck Observatory, which is a private 501(c)3 non-profit organization operated as a scientific partnership among the California Institute of Technology, the University of California, and the National Aeronautics and Space Administration. The Observatory was made possible by the generous financial support of the W. M. Keck Foundation. This research has made use of the Keck Observatory Archive (KOA), which is operated by the W. M. Keck Observatory and the NASA Exoplanet Science Institute (NExScI), under contract with the National Aeronautics and Space Administration.
	
The LBT is an international collaboration among institutions in the United States, Italy, and Germany. LBT Corporation partners are as follows: The University of Arizona on behalf of the Arizona Board of Regents; Istituto Nazionale di Astrofisica, Italy; LBT Beteiligungsgesellschaft, Germany, representing the Max-Planck Society, The Leibniz Institute for Astrophysics Potsdam, and Heidelberg University; The Ohio State University, representing OSU, University of Notre Dame, University of Minnesota and University of Virginia.
	
\bibliography{sun26_solar}{}
\bibliographystyle{aa}

\begin{appendix}
\onecolumn
\section{Equivalent widths and final abundance}
	
Table~\ref{tab:EW} lists the equivalent widths (EWs) measured from the Keck/HIRES, LBT/PEPSI, and Subaru/HDS spectra for the 52 new targets, together with the adopted line list and atomic data, while Table \ref{tab:abund} presents their final differential abundances and associated uncertainties. The complete tables are available in machine-readable format.
	
	\begin{table}
		\centering
		\caption{Equivalent widths measured from Keck/HIRES, LBT/PEPSI, and Subaru/HDS spectra\label{tab:EW}}
		\resizebox{\textwidth}{!}{%
			\begin{tabular}{cccccccccccccc}
				\hline\hline
				$\lambda^a$ & Ion & ExPot & log$gf$ & Damp &
				Sun$^b$ & TOI-1723 & TOI-1799 & TOI-1742 & TOI-1710 & TOI-1691 & TOI-1473 & TOI-1136 & ... \\
				(\AA) &  & (eV) &  &  &
				(m\AA) & (m\AA) & (m\AA) & (m\AA) & (m\AA) & (m\AA) & (m\AA) & (m\AA) & ... \\
				\hline
				5044.211 & 26.0 & 2.851 & -2.058 & 2.71E-31 & 73.3 & 77.9 & 72.8 & 80.4 & 77.7 & 76.9 & 75.8 & 77.4 & ... \\
				5054.642 & 26.0 & 3.640 & -1.921 & 4.68E-32 & 40.5 & 46.5 & 44.7 & 49.2 & 45.2 & 44.3 & 43.1 & 39.7 & ... \\
				5127.359 & 26.0 & 0.915 & -3.307 & 1.84E-32 & 96.0 & 100.2 & 97.6 & 104.0 & 101.0 & 99.3 & 98.5 & 100.8 & ... \\
				5127.679 & 26.0 & 0.052 & -6.125 & 1.20E-32 & 19.1 & 21.9 & 22.0 & 23.7 & 22.2 & 21.7 & 20.9 & 11.8 & ... \\
				5150.839 & 26.0 & 0.989 & -3.003 & 3.45E-32 & 89.3 & 94.8 & 91.0 & 98.6 & 94.7 & 93.5 & 92.1 & 93.8 & ... \\
				5198.711 & 26.0 & 2.223 & -2.135 & 6.22E-32 & 62.5 & 66.3 & 64.2 & 69.0 & 66.1 & 65.4 & 64.0 & 65.8 & ... \\
				5225.526 & 26.0 & 0.110 & -4.789 & 9.10E-33 & 79.6 & 83.4 & 81.2 & 87.1 & 84.0 & 83.1 & 82.0 & 83.5 & ... \\
				5242.491 & 26.0 & 3.634 & -0.967 & 1.17E-31 & 71.2 & 75.9 & 73.5 & 78.6 & 76.0 & 75.2 & 74.1 & 75.4 & ... \\
				$\cdots$ & $\cdots$ & $\cdots$ & $\cdots$ & $\cdots$ & $\cdots$ & $\cdots$ & $\cdots$ & $\cdots$ & $\cdots$ & $\cdots$ & $\cdots$ & $\cdots$ & $\cdots$ \\
				5197.576 & 26.1 & 3.230 & -2.348 & 3.12E-32 & 45.2 & 48.0 & 47.1 & 49.8 & 48.3 & 47.6 & 46.9 & 47.5 & ... \\
				5234.625 & 26.1 & 3.221 & -2.279 & 2.89E-32 & 52.1 & 55.0 & 54.2 & 57.3 & 55.8 & 55.0 & 54.3 & 55.2 & ... \\
				$\cdots$ & $\cdots$ & $\cdots$ & $\cdots$ & $\cdots$ & $\cdots$ & $\cdots$ & $\cdots$ & $\cdots$ & $\cdots$ & $\cdots$ & $\cdots$ & $\cdots$ & $\cdots$ \\
				5380.337 & 6.0 & 7.685 & -1.615 & 1.05E-31 & 26.4 & 28.1 & 27.6 & 29.2 & 28.5 & 28.0 & 27.4 & 28.3 & ... \\
				$\cdots$ & $\cdots$ & $\cdots$ & $\cdots$ & $\cdots$ & $\cdots$ & $\cdots$ & $\cdots$ & $\cdots$ & $\cdots$ & $\cdots$ & $\cdots$ & $\cdots$ & $\cdots$ \\
				6154.226 & 11.0 & 2.102 & -1.547 & 2.11E-32 & 38.5 & 41.2 & 40.3 & 42.5 & 41.5 & 40.9 & 40.2 & 41.0 & ... \\
				$\cdots$ & $\cdots$ & $\cdots$ & $\cdots$ & $\cdots$ & $\cdots$ & $\cdots$ & $\cdots$ & $\cdots$ & $\cdots$ & $\cdots$ & $\cdots$ & $\cdots$ & $\cdots$ \\
				6155.134 & 14.0 & 5.619 & -0.786 & 3.77E-32 & 31.2 & 33.0 & 32.5 & 34.1 & 33.4 & 32.9 & 32.3 & 33.1 & ... \\
				6237.319 & 14.0 & 5.614 & -1.030 & 3.20E-32 & 25.7 & 27.5 & 27.0 & 28.4 & 27.8 & 27.2 & 26.8 & 27.6 & ... \\
				$\cdots$ & $\cdots$ & $\cdots$ & $\cdots$ & $\cdots$ & $\cdots$ & $\cdots$ & $\cdots$ & $\cdots$ & $\cdots$ & $\cdots$ & $\cdots$ & $\cdots$ & $\cdots$ \\
				\hline
			\end{tabular}%
		}
		\par\smallskip
		\footnotesize
		\noindent\textit{Notes.}
		$^a$ Columns 1--5 list the atomic data for each spectral line, including the wavelength in angstroms (\AA), the ion identifier, the excitation potential, the logarithm of the oscillator strength, and the damping constant. The ion identifier follows the standard convention in which the integer part specifies the atomic number and the decimal part indicates the ionization stage, with 0 corresponding to neutral species and 1 to singly ionized species. For example, 26.0 denotes Fe~I, 26.1 denotes Fe~II, and 6.0 denotes C~I. \\
		$^b$ Column 6 gives the solar equivalent widths measured from the Keck/HIRES spectrum. Columns 7--24 list the equivalent widths (EWs) measured for the 18 targets observed with Keck/HIRES, Columns 25--57 those for the 33 targets observed with LBT/PEPSI, and Columns 58--61 those for the four targets observed with Subaru/HDS. \\
		A portion of the table is shown here; the full table is available in machine-readable format.
	\end{table}
	
	\begin{table}
		\centering
		\caption{Final stellar abundances\label{tab:abund}}
		\resizebox{\textwidth}{!}{%
			\begin{tabular}{lcccccccccccccc}
				\hline\hline
				Species$^a$ & Atom$^a$ &
				\multicolumn{4}{c}{TOI-1723$^b$} &
				\multicolumn{4}{c}{TOI-1799$^b$} &
				\multicolumn{4}{c}{TOI-1742$^b$} & ... \\
				& &
				[X/H] & $\sigma_\mu$ & err$_{\rm atm}$ & err$_{\rm comb}$ &
				[X/H] & $\sigma_\mu$ & err$_{\rm atm}$ & err$_{\rm comb}$ &
				[X/H] & $\sigma_\mu$ & err$_{\rm atm}$ & err$_{\rm comb}$ & ... \\
				\hline
				Fe & 26 & 0.070 & 0.004 & 0.013 & 0.014 & -0.011 & 0.004 & 0.014 & 0.015 & 0.158 & 0.003 & 0.012 & 0.012 & ... \\
				C  & 6  & 0.117 & 0.035 & 0.013 & 0.037 & 0.036 & 0.030 & 0.029 & 0.042 & 0.153 & 0.010 & 0.016 & 0.019 & ... \\
				O  & 8  & 0.004 & 0.017 & 0.017 & 0.024 & 0.021 & 0.042 & 0.017 & 0.045 & 0.163 & 0.025 & 0.020 & 0.032 & ... \\
				... & ... & ... & ... & ... & ... & ... & ... & ... & ... & ... & ... & ... & ... & ... \\
				S  & 16 & 0.030 & 0.015 & 0.011 & 0.019 & 0.079 & 0.030 & 0.012 & 0.032 & 0.189 & 0.053 & 0.016 & 0.055 & ... \\
				K  & 19 & 0.013 & 0.010 & 0.021 & 0.023 & 0.009 & 0.010 & 0.010 & 0.014 & 0.157 & 0.008 & 0.019 & 0.021 & ... \\
				Ca & 20 & 0.053 & 0.008 & 0.013 & 0.015 & -0.013 & 0.013 & 0.007 & 0.015 & 0.144 & 0.008 & 0.008 & 0.011 & ... \\
				Sc & 21 & 0.053 & 0.009 & 0.014 & 0.017 & 0.013 & 0.016 & 0.033 & 0.037 & 0.205 & 0.023 & 0.014 & 0.027 & ... \\
				... & ... & ... & ... & ... & ... & ... & ... & ... & ... & ... & ... & ... & ... & ... \\
				Cu & 29 & 0.112 & 0.022 & 0.009 & 0.024 & 0.054 & 0.003 & 0.008 & 0.009 & 0.266 & 0.015 & 0.007 & 0.017 & ... \\
				Zn & 30 & 0.051 & 0.006 & 0.008 & 0.010 & -0.002 & 0.022 & 0.009 & 0.024 & 0.294 & 0.072 & 0.007 & 0.072 & ... \\
				... & ... & ... & ... & ... & ... & ... & ... & ... & ... & ... & ... & ... & ... & ... \\
				\hline
			\end{tabular}%
		}
		\par\smallskip
		\footnotesize
		\noindent\textit{Notes.}
		$^a$ Species and atomic number. The oxygen abundances have been corrected for non-LTE effects. \\
		$^b$ For each star, we list the mean abundance of each element ([X/H]), the standard error of the mean ($\sigma_\mu$), the uncertainty propagated from stellar atmospheric parameters, and the total uncertainty, computed by adding these contributions in quadrature. The first three stars from Keck/HIRES are shown here as an example. The full table includes abundances for 18 stars observed with Keck/HIRES, followed by 33 stars observed with LBT/PEPSI and four stars observed with Subaru/HDS. The full table is available in machine-readable format.
	\end{table}
	
\end{appendix}

\end{document}